# A Location-based Game for Two Generations: Teaching Mobile Technology to the Elderly with the Support of Young Volunteers

Wiesław Kopeć[1], Katarzyna Abramczuk[2], Bartłomiej Balcerzak[1], Marta Juźwin[1], Katarzyna Gniadzik[1], Grzegorz Kowalik[1], and Radosław Nielek[1]

[1] Polish-Japanese Academy of Information Technology,
Koszykowa 86, 02-008 Warsaw, Poland
`{wieslaw.kopec, bartlomiej.balcerzak, marta.juzwin,`
`katarzyna.gniadzik, grzegorz.kowalik, nielek}@pja.edu.pl`
[2] Institute of Sociology, University of Warsaw,
Karowa 18, 00-927 Warsaw, Poland
`k.abramczuk@uw.edu.pl`

**Abstract.** This paper presents a cooperative location-based game for the elderly with the use of tablets equipped with mobile application. The game was designed to tackle at once several crucial topics related to the issue of aging, namely the social inclusion, education in the field of modern technology, motivation for learning as well as physical activity. Mixed-aged teams consisting of two players: a junior and a senior took part in the game. The preliminary results suggest that the game can successfully address a number of issues including improving the elderly technical skills, increasing the elderly physical activity as well as positive intergenerational interaction. The paper describes the game setup in details and presents some initial data gathered during the gameplay.

## 1 Introduction

The process of aging accelerates in European societies and becomes one of the greatest social and economic challenges. Specialists in the field of demographics predict that by 2050 over a quarter of the EU population will be composed of people that will be 65+. Therefore, it is of the highest importance for us to learn how to include this demographic in various activities, facilitate their participation in modern technology-based world, and ultimately apply their strength for the benefit of the society at large. This objective cannot be achieved without solid knowledge about the psychological and sociological processes associated with aging. Modern ways of dealing with the problem are inevitably related to the Information and Communication Technologies (ICT).

The location-based-game research case described in this paper is a part of the project called Living Laboratory (LivingLab) initialized, developed and implemented at Polish-Japanese Academy of Information Technology (PJAIT) located in Warsaw, Poland. It is run in cooperation with the Municipality of Warsaw. Its goals are related to the vital problems of social informatics and are connected with a broad range of GOWELL topics including research and development of solutions for active aging and healthy living, game application for better lifestyle and well-being, positive gaming, stress management and technologies enhancing social well-being. Currently it is in testing stage, with over 200 hundred elderly participants, most of whom are seniors who completed the basic computer course provided by the City of Warsaw.

In this paper we describe one of the LivingLab PJAIT attempts to tackle at once several crucial topics related to the aging issue, namely improve social inclusion of the elderly, their technological skills in the field of mobile technologies, motivation for learning as well as physical well-being, due also to a positive intergenerational interaction. The tool that allowed us to combine all those issues was a location-based game "Stroll Around Yesterday" which joined historical knowledge with the use of tablets and interaction in mixed-age teams of two players (a senior and a PJAIT student of the computer science track). We also chose this form of activity to help a IT student to understand the requirements that should be taken into account in the process of creating software applications for senior citizens.

## 2 Related work

The multidisciplinary research approach mentioned above implies the necessity of considering various perspectives related to a broad range of topics including social inclusion of elders, intergenerational interaction and stereotypes, ICT skills and barriers as well as well-being of the elderly. The problem analysis leads us from the motivation and social activities of older adults to the use of location-based games and modern ICT technology.

A general model for changes in social activities at different stages of life was proposed by Carstensen [1], who suggested that limited activity among older adults may be an adaptive mechanism for coping with changing environment. Previously, models of motivation of the elderly were also studied by Vallerand [2], where motivation was divided into measurable mechanisms. Similar measures for apathy among older adults were proposed by Resnick [3]. Motivation has ramification in many fields of social activity. Hence, researchers conducted studies about the interaction between the elderly motivation and their purchase decisions [4], physical and sport activity [5][6], psychological adjustment in nursing homes [7], family interactions [8] and workplace performance [9], [10], volunteering for social activities and leisure participation [11], [12], and, what is particularly important for the scope of our paper, use of internet-based communication and use of computing technologies [13], [14], [15], [16], [17], [18]. A separate aspect of this research is also connected to investigation into the use of ICT resources in medical treatment of older adults [19].

During all of these studies the observation was made that older adults suffer from a decrease of motivation, and the results suggested the active role of the elderly in restoring the lost motivation and activity. This makes reaching out and providing support for the elderly crucial.

Many methods for achieving this objective have been suggested and tested. Casati et al. [20], [21] have built an on-line platform which encompasses various application designed to motivate older adults to participate in various physical activities. They emphasize the importance of social cues, such as communication with other users, and a coherent narrative for improving the overall performance of the older adults.

Combining location-based games and mobile technologies has also been studied. Avouris [22] reviewed 15 location-based games where mobile technologies were applied and described the impact of the technology on the general performance. Kiefer [23] proposed a classification of different designs of location-based games. Intergenerational aspects of location-based games were considered for example by Charness [9], who studied how participation in such activities differs among younger and older adults. The findings suggested that older adults, while more cooperative, were equally motivated as their younger counterparts.

Another important field of scientific research is related to the intergroup relations and intergenerational interactions. In connection with outdoor activities and location-based games they are sometimes referred to as hybrid reality games [24]. There are some studies on the impact of negative stereotypes on the attitude and performance of the elders. They reveal an interference with intergenerational communication [25], show that stereotypes can both impair and enhance older adults' memory [26], and indicate that the direct, personal contact with members of a different age group can be more effective in improving the intergroup relations than indirect contact [27], [28].

## 3   Gameplay

The game setup was inspired by the study of related work and literature supported by a set of best practices conveyed by external consultants experienced in location-based game design and elderly outdoor activities (e.g. city tour guides). The game "Stroll Around Yesterday" joins historical knowledge with the use of tablets and interaction in mixed-age teams of two players: a senior and a junior. The study concept was to stimulate interaction and cooperation between the team partners: on the one hand, the elderly participants were using the device and mobile apps with an indirect assistance of the younger tech-savvy team member, on the other hand the elderly should be more familiar with the historical and cultural context of the game (location descriptions and hints based on the literature and photos from the past).

Since every successful location-based game needs a good storyline with an alluring plot, we developed a story about a mad scientist, dr von Gestern, who had built a machine disturbing the space-time continuum and bringing back buildings from the past, namely from the communist era. The task for each team was to find all locations, close the wormholes and restore the contemporary buildings. The corresponding promotional materials were created (movie trailer, website, press release, Facebook profile etc.) in order to support the recruitment process of game participants. The preliminary expert consultation along with field tests provided valuable insight and led us to refine the scenario before conducting the first research gameplay. The route was simplified and tailored to the capabilities of the elderly. Finally, the game consisted of four stages, with total route of about 2 km length and duration of about 1,5 h including mid-time coffee break.

The initial gameplay "Stroll Around Yesterday" was held in Warsaw in the area of the Constitution Square (Southern part of the city center) on October 4[th] 2015. It was during Warsaw Senior Week, as a part the of the local International Day of Older Persons celebration (UN established, observed on October 1[st]).

The teams consisted of two people: a senior (the user of the LivingLab PJAIT platform) and a junior (PJAIT student of the computer science track). They were all equipped with space-time fixing modules i.e. tablets provided by the LivingLab team with preloaded software including the special game application.

The tablets could only be operated by the older team member. Directions to the next location were provided by our mobile application. The application was displaying current GPS position on the map alongside with the additional destination hints: textual (based on literature and cultural context) and visual (old B&W depiction gradually transforming into contemporary colorful location photograph). Having reached the location players were obliged to close a wormhole by completing certain task, which always included using the tablet. The tasks were connected to the storyline and related to various activities usually performed on mobile devices. On the first location the task was to connect to the Wi-Fi hotspot and scan the QR code. On the second station the task was to take a panoramic picture. On the third location players were to search the information on the Internet. The game ended at the starting point with playing a puzzle game on the tablet. The gameplay was accompanied by the pre- and post-game evaluation as well as in-game observation.

The follow-up game (shortened demonstration version) was held on October 9[th] as a subject of a field visit of the "AFE-INNOVNET: Towards an Age-Friendly Europe" international workshop.

## 4   Results

The game was played by 30 participants organized in 15 two-person teams. The general impression was very positive. All participants enjoyed the event and we received many requests for continuation. The older people were satisfied with their performance and pleased with cooperation with the junior counterparts. Both groups claimed that during the game there was a true cooperation were both sides had an opportunity for an initiative and contributed nearly equally to the success of the team (all the teams completed the game).

Below we present some preliminary statistics obtained in surveys that were deployed before and after the gameplay. Due to limited space we only present the most basic results. First, we describe shortly the elderly participants' profile. Then we move on to the performance evaluation. Finally we are signaling some findings concerning intergroup relations.

An average senior player was almost seventy-year-old retired woman from a large city (73% female participants, average age 69, youngest 60, oldest 86), rather well educated (60% of higher education) with basic computer skills and motivation to learn how to use tablets. She uses smartphones rather regularly (60%) and considers mobile devices very useful (75% before the game, 87% after the game), but has a limited knowledge about using it (40% of younger participants evaluated their older counterpart as having little knowledge about how to use tablet; some participants claimed that it was the first time they used the tablet, nevertheless they managed to complete the game). From additional surveys performed on the LivingLab platform we also know that our senior is rather independent: 50% live on their own, without family members, 75% have a PC, and 63% use it without any assistance. They have a broad variety of interests from cooking and crosswords to chemometrics and fitness.

To evaluate the performance, we asked both groups (seniors and juniors) to choose the most accurate description of what had happened in each game task on a 5-point scale from "junior completed the task" to "senior completed the task without any assistance". The reports were rather consistent within most pairs. In very few cases the evaluations by two parties differed by more than one category, which demonstrates their reliability. According to the participants' evaluations in most cases the senior completed the tasks instructed by her/his partner. The most problematic task was establishing Wi-Fi connection where oftentimes-direct junior assistance was needed.

We asked both parties to indicate which game task in their view could be performed again by senior without any assistance. The results are presented in Figure 1. An interesting conclusion is that seniors generally underestimate their performance in comparison with the external estimate of their capabilities. This refers particularly to those tasks that are relatively unfamiliar to their experience such as taking a panoramic photo, scanning a QR code or playing a mobile game. On the other hand, seniors overestimate their competence in more common tasks such as establishing a Wi-Fi connection or searching the Web.

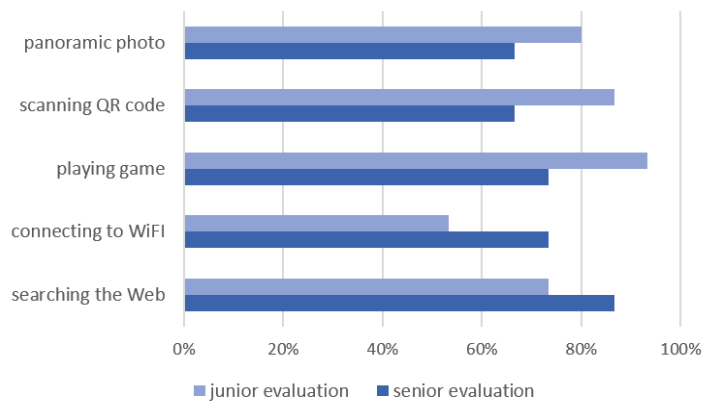

**Fig. 1.** The game tasks evaluation by both groups of participants indicating which tasks could be performed again by senior without any assistance.

According to the contact theory [29] [30], if certain conditions are met, contact between different groups can improve their attitudes towards each other. As our game fulfilled most of the criteria for a successful contact, we decided to explore whether we can observe some shift in attitudes. We asked both seniors and juniors to evaluate members of the other age group with the use of a set of 12 antonyms selected from Aging Semantic Differential by Rozencranz and McNevin [31]. The list of the antonyms can be found in Table 1.

**Table 1.** List of 12 dimensions for intergroup evaluation.

| Positive | Negative |
|---|---|
| productive | unproductive |
| active | passive |
| aggressive | defensive |
| independent | dependent |
| organized | disorganized |
| decisive | indecisive |
| cooperative | uncooperative |
| flexible | inflexible |
| hopeful | dejected |
| trustful | suspicious |
| pleasant | unpleasant |
| exciting | dull |

Before and after the game the participants were asked to evaluate a general other (some unspecified member of the other age group) using the antonym pairs. The chart below shows the median evaluations of the general other from the other age group for junior and senior evaluators.

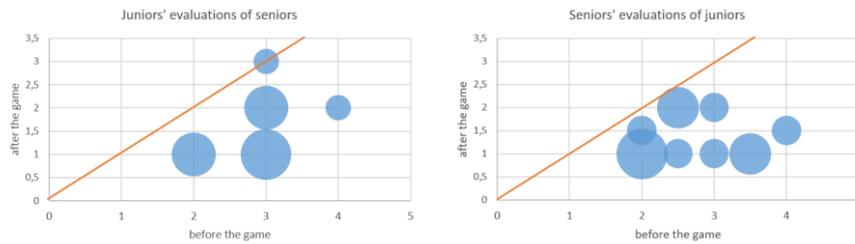

**Fig. 2.** Median evaluations of general other before and after the game.

The larger values indicate less favorable opinions. The size of the circles matches the number of antonyms with the corresponding coordinates. For example, the outermost left circle on the left hand side graph corresponds to 3 antonyms for which the median evaluations of the general older other was 2 before the game and 1 after the game. The straight lines indicate equal medians. This is where all the circles would lie, was there no shift in attitudes. As virtually all of the circles in Figure 2 are below that line, we can hypothesize that for almost all the dimensions the perception of the outgroup improved for both juniors and seniors. A more detailed statistical analysis is needed to validate this claim.

Summing up, we have reasons to believe that this kind of activity is a very promising tool for improving intergroup relations, and diminishing age-based stereotypes. We plan to conduct additional gameplays to verify this result.

## 5  Summary and future work

The game described in this paper proved to be promising in several ways. It can increase the technical skills of the elderly, improve their physical activity and enhance positive intergenerational interaction.

Apart from the scientific results, the outcome from the initial gameplay was lots of hands-on experience. Unfortunately the initial game setup was very demanding and turned out to be resource intensive and time consuming. On the whole, there were dozens of people engaged from actors and movie crew to software developers, testers and gameplay staff.

In the nearest future we plan to retake the original game to verify the outcome, as well as to continue the struggle in order to prepare more concise and robust setup of the game which could be replayed automatically by a larger group of participants without the need of relying on the human staff, since reusability and scalability are crucial for various research approaches, i.e. crowdsourcing. We also plan to develop a setup that could be used indoors with the use of QR codes and BT beacons.

**Acknowledgments.** This project has received funding from the European Union's Horizon 2020 Research and Innovation Programme under the Marie Skłodowska-Curie grant agreement No 690962.